\documentclass[12pt]{article}
\usepackage{graphicx}

\newcommand{\beqn}{\begin{equation}}
\newcommand{\eeqn}{\end{equation}}
\newcommand{\bcen}{\begin{center}}
\newcommand{\ecen}{\end{center}}
\newcommand{\bfig}{\begin{figure}[!htbp]}
\newcommand{\efig}{\end{figure}}

\newcommand{\asotpi}{\frac{\alpha_s}{2\pi}}
\newcommand{\aemotpi}{\frac{\alpha_{em}}{2\pi}}
\newcommand{\aemofpi}{\frac{\alpha_{em}}{4\pi}}
\newcommand{\ftg}{F_2^{\gamma}}

\catcode`\@=11  
\@addtoreset{equation}{section}
\def\theequation{\arabic{section}.\arabic{equation}}

\begin{document}
\begin{titlepage}
\begin{flushright}
hep-ph/0008275
\end{flushright}
\bcen
      {
      \Large{
            {\bf
                  Photon Structure Functions

                  \smallskip

                  \ \ beyond the SUSY Threshold
            }
            }
      }
\ecen

\bigskip
\bigskip
\bcen

      D. A. Ross \ and \ L. J. Weston

\ecen

\bigskip
\bcen
      {\it
            Department of Physics \& Astronomy, University of Southampton,
            Southampton S017 1BJ, UK
      }
\ecen

\bigskip
\bigskip
\bigskip
\bcen
      {\bf Abstract}
\ecen

We evolve virtual photon parton densities up
to the SUSY threshold and higher using coupled inhomogeneous
DGLAP differential equations.
Reliable input parameterizations were available from the c-quark threshold.
Limited $P^2$ ( target photon virtuality ) dependence is observed.
The difference to the photon structure function is shown to be
significant with the introduction of SUSY dependent splitting
functions.  A negligible difference is observed by letting the
gluino mass enter after the squark mass.  An effort is made to
include the squark threshold effect in such a way that both the renormalization
group equations are satisfied and the perturbative calculation is
reproduced.

\end{titlepage}
\section{Introduction}

There has recently been a great deal of interest in the structure
function of the photon. This is obtained from the scattering
cross-section between a highly virtual photon with large square momentum
$Q^2$ used as a probe and a nearly on-shell
target photon with square momentum
$P^2,\ (P^2 \ll Q^2)$. If the target square momentum is also large
(whilst maintaining the inequality $P^2 \ll Q^2$), the entire structure
function can be calculated using renormalization group improved perturbation
theory, whereas for low $P^2$ one is limited to a determination
of the $Q^2$ dependence and, as in the case of deep-inelastic
electron-proton scattering, one needs input information
on the structure functions at some (low) value of $Q^2$,
which cannot be determined by perturbation theory.
 A study of the photon structure function as a function of $P^2$
therefore provides information on the extrapolation between
the perturbative and non-perturbative regimes of QCD.

Heuristically, one talks about separating the structure function
into ``direct'' and ``resolved'' contributions. The former being
exactly calculable in perturbation theory and the latter involving
the uncalculable probability that the photon splits into other fundamental
particles before being probed. Whereas such a picture is useful
at the leading order level, higher order corrections mix these contributions.
A formal and more precise analysis was first proposed by Witten \cite{witten}
who pointed out that in an operator product expansion
for  photon-photon scattering the set of operators used in the case
of photon-proton scattering must be augmented by a tower of
 photonic operators, whose matrix elements with the target photon are of
order unity. In the (more intuitive) DGLAP approach one argues
that since the probability of finding a particle other than a photon
inside a photon is of order $\alpha_{em}$, the probability of
finding a ``photon inside a photon'' is unity plus corrections
of order $\alpha_{em}$. The DGLAP analysis must then be augmented
by  further off-diagonal splitting functions $K_q$ and $K_G$
which are interpreted as the perturbative probability for a
photon to emit quark or gluon with a given fraction of its
 longitudinal momentum.

Interest in photon structure functions has recently been rekindled
by the prospect of a future high-energy electron-positron collider
with centre-of-mass energy of up to 1 TeV. Such a machine would enable
an investigation of the photon structure function over a sufficiently
wide range of $Q^2$ and $P^2$ to provide a stringent test of the
evolution of these structure functions. Moreover, if the postulated existence
of Supersymmetry (SUSY) turns out to be vindicated, these structure functions
will reflect the existence of supersymmetric partners within photons.
The contribution to the structure function due to the crossing
of the threshold for the production of squarks, was first calculated
by Reya \cite{TH3504}. However, a consistent analysis of the effect
of supersymmetry on the photon structure function requires a full analysis
of the enlarged DGLAP formalism in which above the SUSY
threshold the standard splitting functions
are augmented with splitting functions involving squarks and gluinos.
This paper reports on such an analysis and displays results in which
it can be seen that SUSY gives rise to a measurable increase in the
$Q^2$ evolution of the structure photon structure function above threshold.
Care must be taken to ensure a consistent treatment of the threshold behaviour
for squark production as one passes through the threshold and this is
discussed in detail.

The outline of the paper is as follows: In Section 2 we discuss the formalism,
outlining the extension of the evolution equations to the regime in which
squarks and gluinos are excited. We also give a description of the threshold
treatment. Section 3 displays our  results obtained from numerical
solution of the enlarged evolution equations. We show the dependences on
the SUSY threshold and on the square momentum $P^2$ of the target photon as well
as on the usual variables $Q^2$ and Bjorken-$x$. In Section 5 we discuss
our conclusions.

\section{Formalism}

We follow the formalism of Gl\"{u}ck and Reya \cite{D282749}.
We will initially be concerned with quark and gluon distributions up to the SUSY
threshold.

The nonsinglet ($T$) and singlet ($\Sigma$) sectors are treated separately,

\begin{eqnarray*}
  T3 & = & 2(u-d) \\
  T8 & = & 2(u+d-2s) \\
  T15 & = & 2(u+d+s-3c) \\
  T24 & = & 2(u+d+s+c-4b) \\
  T35 & = & 2(u+d+s+c+b-5t) \\
  \Sigma & = & 2 \, \sum_{i}^{f} \, q_i
\end{eqnarray*}

\noindent where $u,d,s,c,b,t$ refer to the relevant quark
distributions.  The factor of $2$ accounts for the anti-quark distribution
since for a photon $q_i = \bar{q}_i$.   $f$ runs
over all relevant quark flavours. Each
quark distribution is zero at and below its threshold hence each new
non-singlet ($T$) is equal to the singlet ($\Sigma$) at threshold.

The evolution is controlled by the following inhomogeneous DGLAP
differential equations,

\beqn \label{A}
\frac{d T }{d \, \ln Q^2} \, = \, P_{T T} \otimes
T \, + \, K_{T}
\eeqn

\noindent for each singlet ($T$) and the coupled equations,

\[
\frac{d \Sigma}{d \, \ln Q^2} \, = \, P_{\Sigma \Sigma} \otimes
\Sigma \, + \, P_{\Sigma G} \otimes G \, + \, K_{\Sigma}
\]

\beqn \label{B}
\frac{dG}{d \, \ln Q^2} \, = \, P_{G \Sigma} \otimes \Sigma  \, + \,
P_{G G} \otimes G \, + \, K_{G}
\eeqn

\noindent for the singlet ($\Sigma$) and gluon ($G$) sector.

For each distribution $F(x,Q^2)$ above, the convolution $\otimes$ is
defined as,

\begin{equation}\label{C}
  P_{ij} \otimes F_{j} \ \equiv
      \ \int_x^1 \, \frac{dy}{y} \, P_{ij}\left(\frac{x}{y},Q^2\right) F_{j}(y,Q^2)
\end{equation}

\noindent where $P_{ij}(x,Q^2)$ consists of the splitting functions
$p_{ij}^{(0)}$ in Leading Order (LO) and $p_{ij}^{(1)}$ in next to leading order (NLO),

\begin{equation}\label{D}
  P_{ij} = \left[ \asotpi \right] \, p_{ij}^{(0)}
         + \left[ \asotpi \right]^2 p_{ij}^{(1)} + \cdots
\end{equation}

The main difference between the evolution of the photon structure function
and that of the proton structure function is the presence of the inhomogeneous
$K_i$ terms in the evolution equations.
Essentially these consist of $\gamma \to$ quark and $\gamma \to$ gluon splitting functions.
They appear in
the evolution equations without any convolution with a parton distribution
since the ``photon density'' inside a photon is taken to
 be $\delta(1-x)$ up to corrections of order $\alpha_{em}$.

\begin{equation}\label{E}
  K_{i} = \left[ \aemotpi \right] \, k_{i}^{(0)}
    +  \left[ \aemotpi \right] \left[ \asotpi \right] \, k_{i}^{(1)} + \cdots
\end{equation}

$F_2^\gamma$ in LO is given by,

\begin{equation}\label{F}
\frac{1}{x} F_2^{\gamma} \, = \, \left\{ q_{\mbox{\scriptsize{NS}}} \, + \, \langle e^2
\rangle \, \Sigma \right\}
\end{equation}

\noindent where

\[
q_{\mbox{\scriptsize{NS}}} \, = \, \sum_f \left( e_q^2 - \langle
e^2 \rangle \right) \left( q_i + \bar{q}_i \right) \ \ \ , \ \ \
\langle e^k \rangle \, = \, \frac{1}{f} \sum_f e_q^k
\]

\noindent $\alpha_s(Q^2)$ evolves according to

\begin{equation}\label{G}
\frac{\alpha_s(Q^2)}{4 \pi} = \frac{1}{\beta_0 \ln Q^2 / \Lambda^2} -
        \frac{\beta_1}{\beta_0^3}
        \frac{\ln \, (\ln Q^2 / \Lambda^2)}{(\ln Q^2 / \Lambda^2)^2}
\end{equation}

\noindent where $\beta_0 = 11 - 2f/3$ and $\beta_1 = 102 - 38f/3$.
All expressions refer to the $\overline{M\!S}$ renormalization scheme
hence we use $\Lambda_{\overline{M\!S}}$ which depends $f$.
We evolve in NLO to the t-quark threshold and then we evolve in LO thereafter.
This is because we can only evolve in LO above the SUSY threshold so in order
to compare $F_2^{\gamma}$ for different values of the squark mass $M_s$ we must
evolve in the same way from the t-quark threshold.
Quark masses are taken as $M(c) = 1.5 \, GeV, \ \ M(b) = 4.5 \, GeV, \ \
M(t) = 174 \, GeV$.

\bigskip
Our input data were parameterizations \cite{D545515} of virtual
photon parton densities taken at the c-quark threshold.  A c-quark
mass of $1.5 \, GeV$ limits $P^2$ to less than $1.8 \, GeV$ which
gives a small ratio $r = P^2 / Q^2 \simeq {10}^{-6}$ at high
$Q^2$. We could not find reliable parameterizations at higher
$Q^2$ that were dependent on $P^2$.

\bigskip
In order to evolve to the SUSY threshold we use LO and NLO
splitting functions \cite{ES} in (Eq.~\ref{D}) and
inhomogeneous terms \cite{D282749} in (Eq.~\ref{E}).

\bigskip
Above the SUSY threshold $M_s$ we are also concerned with squark and
gluino distributions. As before we have a nonsinglet (S) and singlet
($\Gamma$) sector,

\begin{eqnarray*}
  S3 & = & 4(S \! u-S \! d) \\
  S8 & = & 4(S \! u+S \! d-2S \! s) \\
  S15 & = & 4(S \! u+S \! d+S \! s-3S \! c) \\
  S24 & = & 4(S \! u+S \! d+S \! s+S \! c-4S \! b) \\
  S35 & = & 4(S \! u+S \! d+S \! s+S \! c+S \! b-5S \! t) \\
  \Gamma & = & 4 \, \sum_{i}^{f} \, S \! \bar{q}_i
\end{eqnarray*}

\noindent where $S \! u,S \! u,S \! s,S \! c,S \! b,S \! t$ refer to the squark
distributions.  The factor of $4$ arises because $S \! q_i^R = S
\! q_i^L = S \! \bar{q}_i^R = S \! \bar{q}_i^L$. For simplicity all squark
distributions start at zero at the SUSY threshold $M_s$, although we
could introduce them in steps as before with the quarks. The
gluino distribution starts at zero at the gluino threshold $M_g$.

The evolution is controlled by the following inhomogeneous DGLAP
differential equations. Each set of nonsinglets are coupled ie.
T3 with S3, T8 with S8, etc...

\[
\frac{d T }{d \, \ln Q^2} \, = \, P_{T T} \otimes T \,
   + \, P_{T S} \otimes S \, + \, K_{T}
\]

\beqn \label{H}
\frac{d S }{d \, \ln Q^2} \, = \, P_{S T} \otimes T \,
   + \, P_{S S} \otimes S \, + \, K_{S}
\eeqn

\noindent Given that in general the gluino mass $M_g$ is greater than
the squark mass $M_s$ the nonsinglet sector evolution is given in the region
$M_g^2 \geq Q^2 \geq M_s^2$ by,

\[
\frac{d \Sigma}{d \, \ln Q^2} \, = \, P_{\Sigma \Sigma} \otimes \Sigma \,
                               + \, P_{\Sigma G} \otimes G \,
                               + \, P_{\Sigma \Gamma} \otimes \Gamma \,
                               + \, K_{\Sigma}
\]

\[
\frac{d G}{d \, \ln Q^2} \, = \, P_{G \Sigma} \otimes \Sigma \,
                               + \, P_{G G} \otimes G \,
                               + \, P_{G \Gamma} \otimes \Gamma \,
                               + \, K_{G}
\]

\beqn \label{I}
\frac{d \Gamma}{d \, \ln Q^2} \, = \, P_{\Gamma \Sigma} \otimes \Sigma \,
                               + \, P_{\Gamma G} \otimes G \,
                               + \, P_{\Gamma \Gamma} \otimes \Gamma \,
                               + \, K_{\Gamma}
\eeqn

\bigskip
\noindent and in the region $Q^2 \geq M_g^2$ by,
\bigskip

\[
\frac{d \Sigma}{d \, \ln Q^2} \, = \, P_{\Sigma \Sigma} \otimes \Sigma \,
                               + \, P_{\Sigma G} \otimes G \,
                               + \, P_{\Sigma \Gamma} \otimes \Gamma \,
                               + \, P_{\Sigma L} \otimes L \,
                               + \, K_{\Sigma}
\]

\[
\frac{d G}{d \, \ln Q^2} \, = \, P_{G \Sigma} \otimes \Sigma \,
                               + \, P_{G G} \otimes G \,
                               + \, P_{G \Gamma} \otimes \Gamma \,
                               + \, P_{G L} \otimes L \,
                               + \, K_{G}
\]

\[
\frac{d \Gamma}{d \, \ln Q^2} \, = \, P_{\Gamma \Sigma} \otimes \Sigma \,
                               + \, P_{\Gamma G} \otimes G \,
                               + \, P_{\Gamma \Gamma} \otimes \Gamma \,
                               + \, P_{\Gamma L} \otimes L \,
                               + \, K_{\Gamma}
\]

\beqn \label{J}
\frac{d L}{d \, \ln Q^2} \, = \, P_{L \Sigma} \otimes \Sigma \,
                               + \, P_{L G} \otimes G \,
                               + \, P_{L \Gamma} \otimes \Gamma \,
                               + \, P_{L L} \otimes L \,
                               + \, K_{L}
\eeqn

\noindent where $L$ is the gluino distribution. In analogy with
the $K_i$ used below the SUSY threshold, $K_\Gamma$ and $K_L$ are the splitting functions
of a photon to a squark and gluino respectively.
In the limit where the gluino mass $M_g$ is taken to be equal to the
squark mass $M_s$ we do not need (Eqs.~\ref{I}).

The $P_{ij}(x,Q^2)$ are now a different SUSY set of LO splitting functions \cite{KR}.
However in order to reproduce the squark threshold condition,

\begin{equation}\label{K}
  Q^2 \frac{(1-x-r \, x)}{x} \ \geq \ 4 M_s^2 
\end{equation}

\noindent we use the SUSY set if this applies or the standard set if it does not.
Importantly this now means that the threshold depends both on $x$ and $Q^2$.
This is possible because a convolution (Eq.~\ref{C}) only feeds each
distribution in the region $y \geq x$.
Hence for a particular $Q^2$, SUSY splitting functions are used below a certain
value of $x$ and standard splitting functions are used above this $x$ during
each convolution.

Our choice of LO inhomogeneous terms (Eq.~\ref{E}) is made in
order to treat the squark threshold in a meaningful way.

The tree level squark contribution to $\ftg$ (Appendix A)
is important in determining the $\gamma \to$ squark
splitting function i.e., the squark inhomogeneous term in
(Eqs.~\ref{H},~\ref{I},~\ref{J}).
The choice is dependent on the squark threshold condition (Eq.~\ref{K}),
which is function of both $x$ and $Q^2$. At a particular $Q^2$ there will
be a region $x > x_s$ where squarks cannot be produced.

The contribution to $F_2^\gamma$ from squark production
( Eq.~\ref{L} )
is obtained from ordinary perturbation theory and includes a term
  proportional to $\ln(Q^2/4M_s^2)$,
which is already accounted for as $K_\Gamma$, by the use of the full SUSY set
of splitting functions discussed above. In order to reproduce the correct
renormalization group improved evolution far above the threshold we subtract
this term (apart from terms proportional to $M_s^2/Q^2$ which
are negligible far above the threshold) from ( Eq.~\ref{L} ) and
introduce the contribution as $S \! B_{\gamma}$ in ( Eq.~\ref{M} ).

Above the SUSY threshold $F_2^\gamma$ is then obtained from
\begin{equation}\label{M}
\frac{1}{x} F_2^{\gamma} \, = \, \left\{ q_{\mbox{\scriptsize{NS}}} \, + \, \langle e^2 \rangle \,
  \Sigma \right\} \, + \, \left\{ S \! q_{\mbox{\scriptsize{NS}}} \, + \,
  \langle e^2 \rangle \, S \Sigma \, \right\} + \,  2 \times 3 f \langle
  e^4 \rangle \left[ \aemofpi \right] S \! B_{\gamma}
\end{equation}

\noindent (the factor of $2$  accounts for right- and left-handed squarks).

The $S \! B_{\gamma}$ term is (up to an overall constant)
the contribution to $F_2^\gamma$ from squark production, with the removal
of the above-mentioned term proportional to $\ln(Q^2/M_s^2)$. i.e.
\begin{equation}\label{N}
S \! B_{\gamma} \ \ \propto \ \ F_{2,squark}^{\gamma} - \ 2x(1-x) \; \ln(Q^2/4M_s^2),
\end{equation}

We note that the difference between using (Eq.~\ref{L})
and (Eq.~\ref{O}) is negligible in our case because
we are limited to $P^2 < 1.8 \, GeV^2$ at the c-quark threshold,
giving an $r \simeq 10^{-6}$ above the SUSY threshold.

However this is a different way of treating the threshold behaviour
from that in \cite{D282749}. At $Q^2 \gg 4M_s^2$ it satisfies the
Renormalization Group equations since the dominant part is in the
inhomogeneous term. In the region $Q^2 \simeq 4M_s^2$ this approach will
reproduce the perturbative calculation with the correct threshold
behaviour up to $(\ln(Q^2/4M_s^2))^2$. There should of course be a small
mismatch at large $Q^2$ and large $x$. However we have eradicated this
by coding the threshold behaviour into each convolution
as a choice of SUSY or non-SUSY splitting functions.
The problem left is that of the discontinuity at the threshold $x$ boundary.
An equivalent problem is discussed in \cite{D453986}
involving the $DIS_{\gamma}$ factorization scheme.
Perturbative instabilities mean we would have to go to NLO and
eliminate the $S \! B_{\gamma}$ term by altering what would be
our NLO inhomogeneous terms (Eq.~\ref{E}).
However NLO SUSY splitting functions are beyond the scope of
the present work.

Finally it should be noted how quickly $\ftg$ changes away from the threshold
with decreasing $x$.
In (Eq.~\ref{L}), the term

\[
v = [ 1 - 4M_s^2x/Q^2(1-x) ]^{1/2}
\]

\noindent moves rapidly away from zero in decreasing x meaning that
the coefficients of

\[
\ln \left( \frac{1+v}{1-v}  \right) \ \ \ {\rm and} \ \ \ v
\]

\noindent in (Eq.~\ref{L}) give rise to a real threshold contained in
the term $S \! B_{\gamma}$ in the perturbatively stable $x$ region.

\bigskip
To summarize, we take parameterizations of quark and gluon distributions
inside a virtual photon at the c-threshold.
Using DGLAP inhomogeneous differential equations we evolve
the relevant non-singlet, singlet and gluon distributions up to the SUSY
threshold. From here we run the distributions separately, including or not, the
effects of squarks and gluinos. At some $\sqrt{Q^2}$ we form
$F_2^{\gamma}$ for the virtual photon in such a way as to take account of the
SUSY threshold condition.

\section{Results}

The variable parameters of the evolution are the $P^2$ (target
virtuality), $M_s$ (squark mass), $M_g$ (gluino mass), $Q^2$
(incident virtuality) and Bjorken $x$. We took these in the ranges,

\[
0 \ \leq \ \sqrt{P^2} \ \leq \ 1300 \, MeV
\]
\[
175 \, GeV \ \leq \ M_s \ \leq \ 300 \, GeV
\]
\[
200 \, GeV \ \leq \ M_g \ \leq \ 300 \, GeV
\]
\[
700 \, GeV \ \leq \ \sqrt{Q^2} \ \leq 1500 \, GeV
\]

\noindent and in all cases $\ftg / \alpha_{em}$ is actually plotted.

\bigskip
Figure~\ref{f1} shows a generalised evolution to $1000 \, GeV$.

The base reference is at $175 \, GeV$.
There is a considerable difference with the SUSY mixing.
Note that allowing the gluino mass to be greater than the
squark mass produces a negligible effect.
Note also that the evolution graphs coincide above the squark threshold
condition (Eq.~\ref{K}), this being due to it being encoded into
each convolution.

The discontinuity at the threshold can be seen.
This results in a sudden drop in $\ftg$ below the control
non-SUSY graph, this being due to the perturbative instabilities
already discussed.

\bigskip
It is important to note the rapidity with which $\ftg$ increases
as $x$ decreases away from the threshold.
This effect due to the $S \! B_{\gamma}$ term in the perturbatively
stable region is discussed in section 2.

\bigskip
Figure~\ref{f2} shows $P^2$ dependence up to $1300 \, MeV$.
From here on we plot $M_g = M_s$ since we have shown the $M_g > M_s$
difference to be negligible.

We are limited by our parameterizations in that they are restricted
in $P^2$ at the c-quark threshold.
However non-negligible differences can be noted in low $x$ even at
$\sqrt{P^2} = 1300 \, MeV$.

\bigskip
Figure~\ref{f3} shows $M_s$ dependence between $175 \, GeV$
and $300 \, GeV$.

The base graph is without the SUSY contributions.
As expected, as the squark mass $M_s$ increases the distribution's
approach to the non-SUSY limit.
Also the thresholds move to lower x as the threshold condition (Eq.~\ref{K})
requires higher $Q^2$ to produce squarks of higher mass.

\bigskip
Figures~\ref{f4} and~\ref{f5} show how $\ftg$ varies with
$\sqrt{Q^2}$ at two fixed values of $x$.
All graphs show how the distributions must approach the non-SUSY
distribution for high $M_s$.
However for $x = 0.66$ we can see the gradual approach to a threshold
in increasing $M_s$.
For $M_s = 275 \, GeV$ it is evident that for low $Q^2$ squarks cannot be
produced and the distribution coincides with the non-SUSY distribution.
Then apart from the perturbative instability the distribution rises in
higher $Q^2$.

\bfig
\bcen
\includegraphics[width = \textwidth,height=15cm]{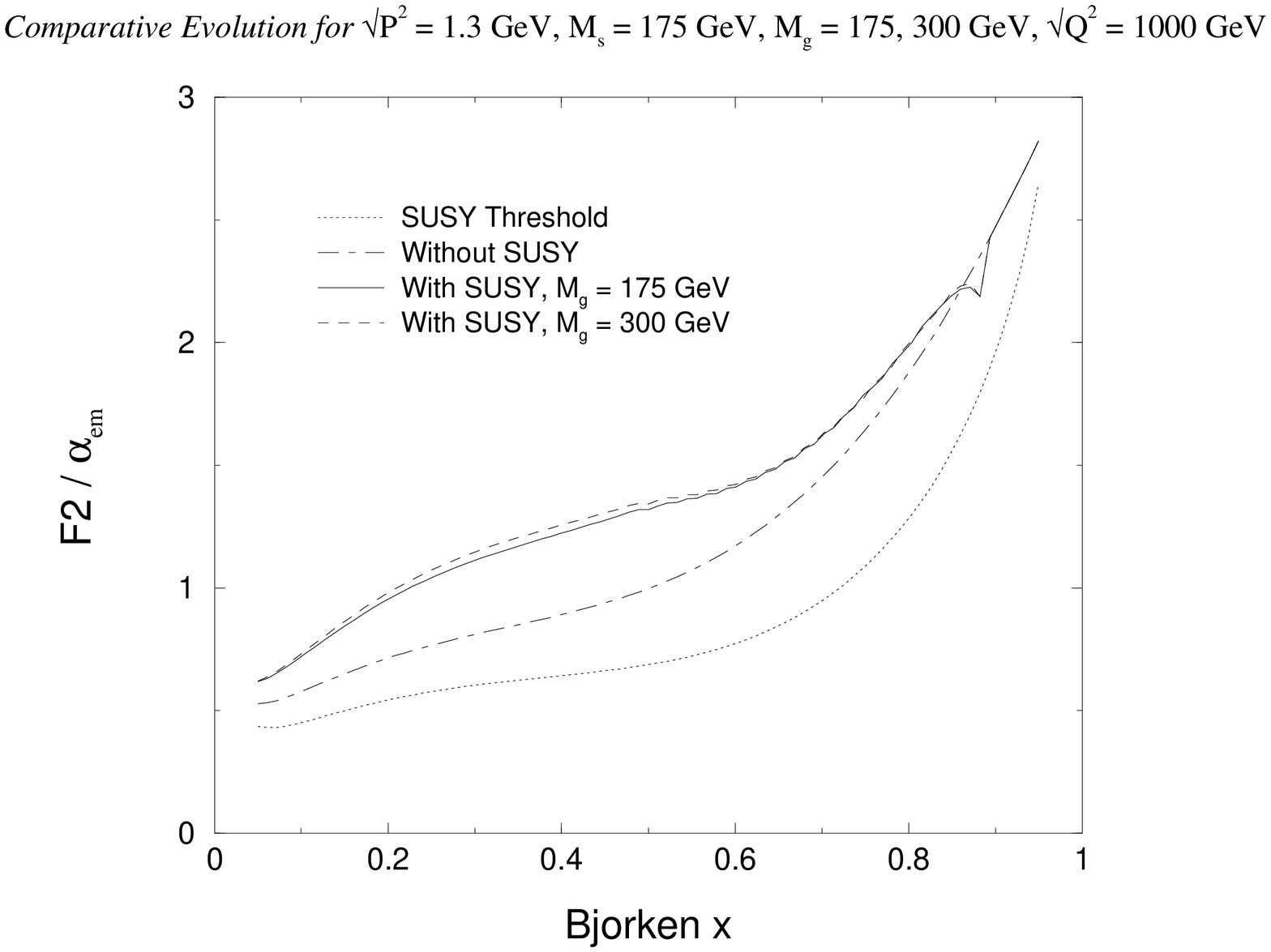}
\ecen
\caption{Comparative Evolution of Structure Function with and without
SUSY splitting functions. Difference due to a higher gluino mass $M_g$
is negligible.}\label{f1}
\efig

\bfig
\bcen
\includegraphics[width= \textwidth,height=15cm]{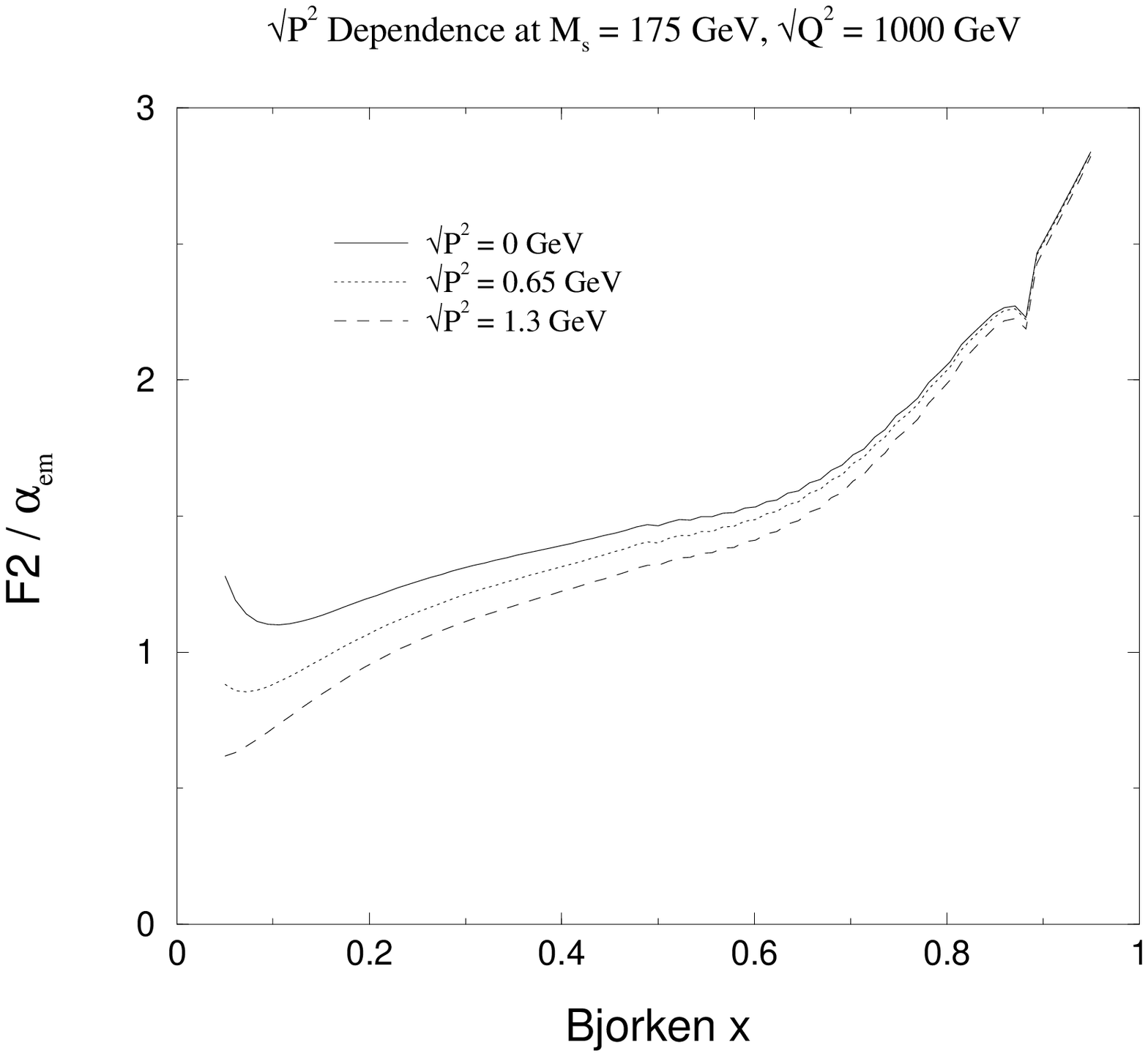}
\ecen
\caption{$\sqrt{P^2}$ dependence of structure function for fixed squark mass $M_s$ at
a fixed probe virtuality $\sqrt{Q^2}$}\label{f2}
\efig

\bfig
\bcen
\includegraphics[width= \textwidth,height=15cm]{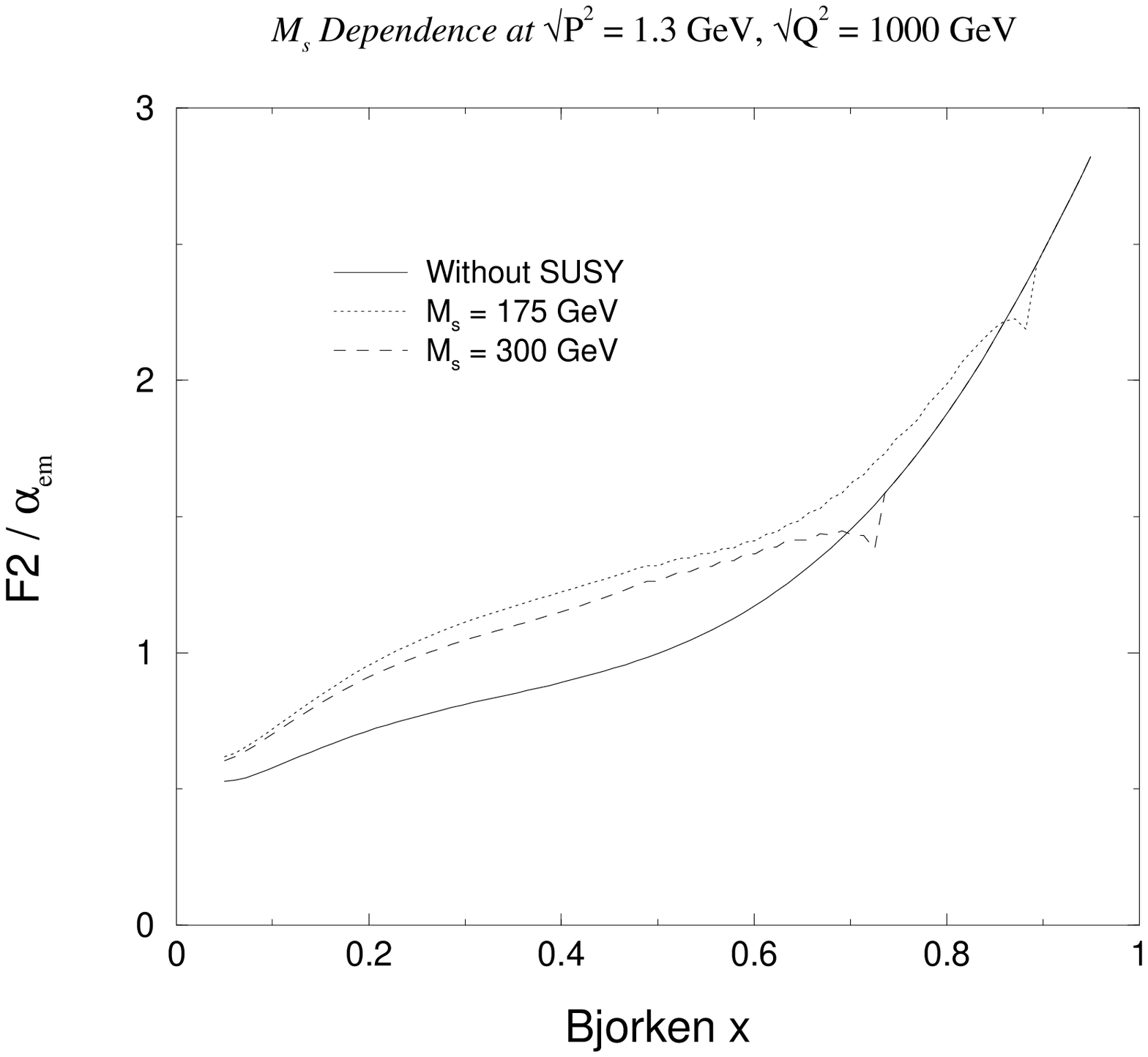}
\ecen
\caption{Dependence of structure function on squark mass $M_s$ at a fixed
target virtuality $\sqrt{P^2}$ and probe virtuality $\sqrt{Q^2}$}\label{f3}
\efig

\bfig
\bcen
\includegraphics[width= \textwidth,height=15cm]{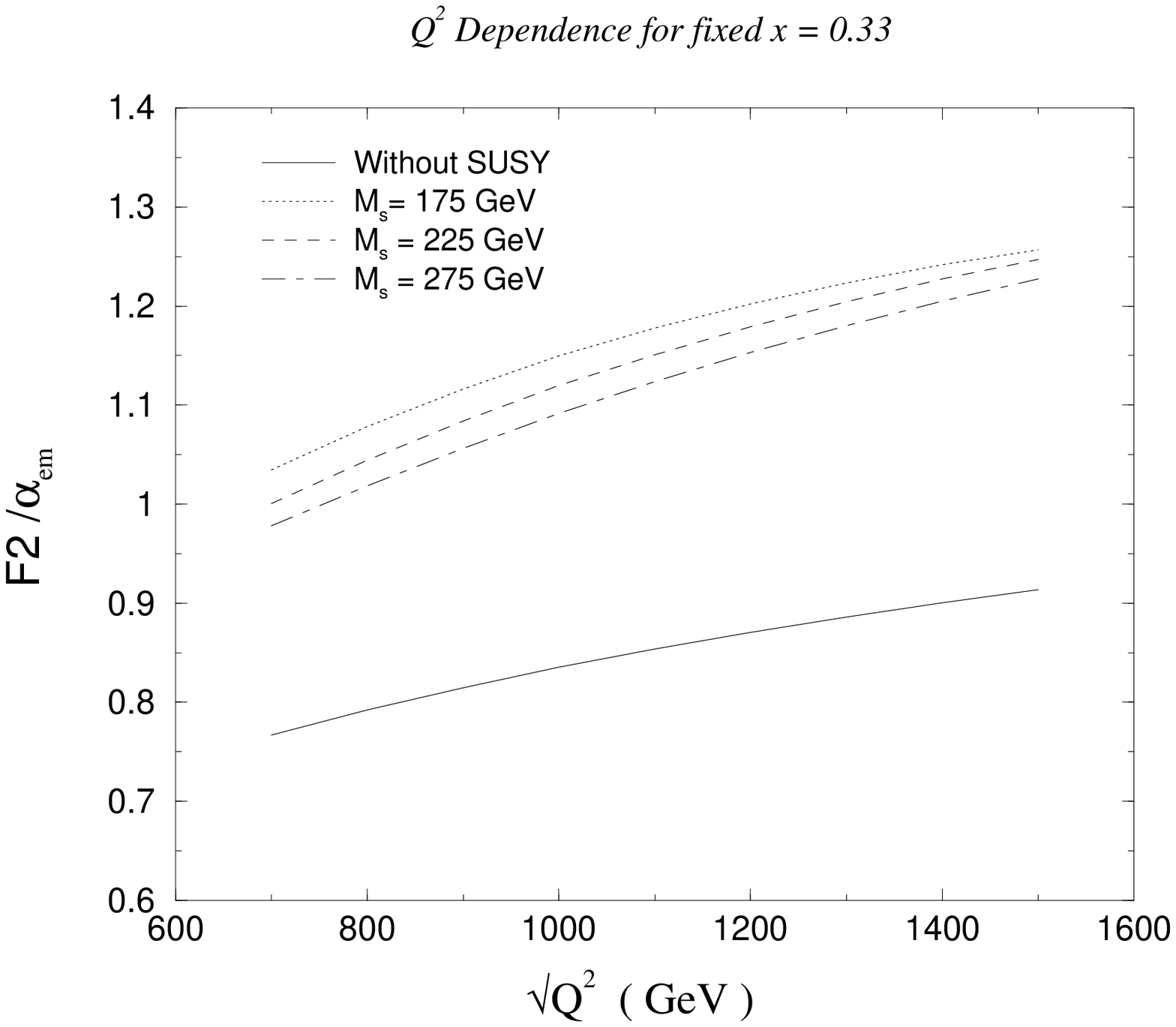}
\ecen
\caption{$x = 0.33$}
\label{f4}
\efig

\bigskip
\bfig
\bcen
\includegraphics[width= \textwidth,height=15cm]{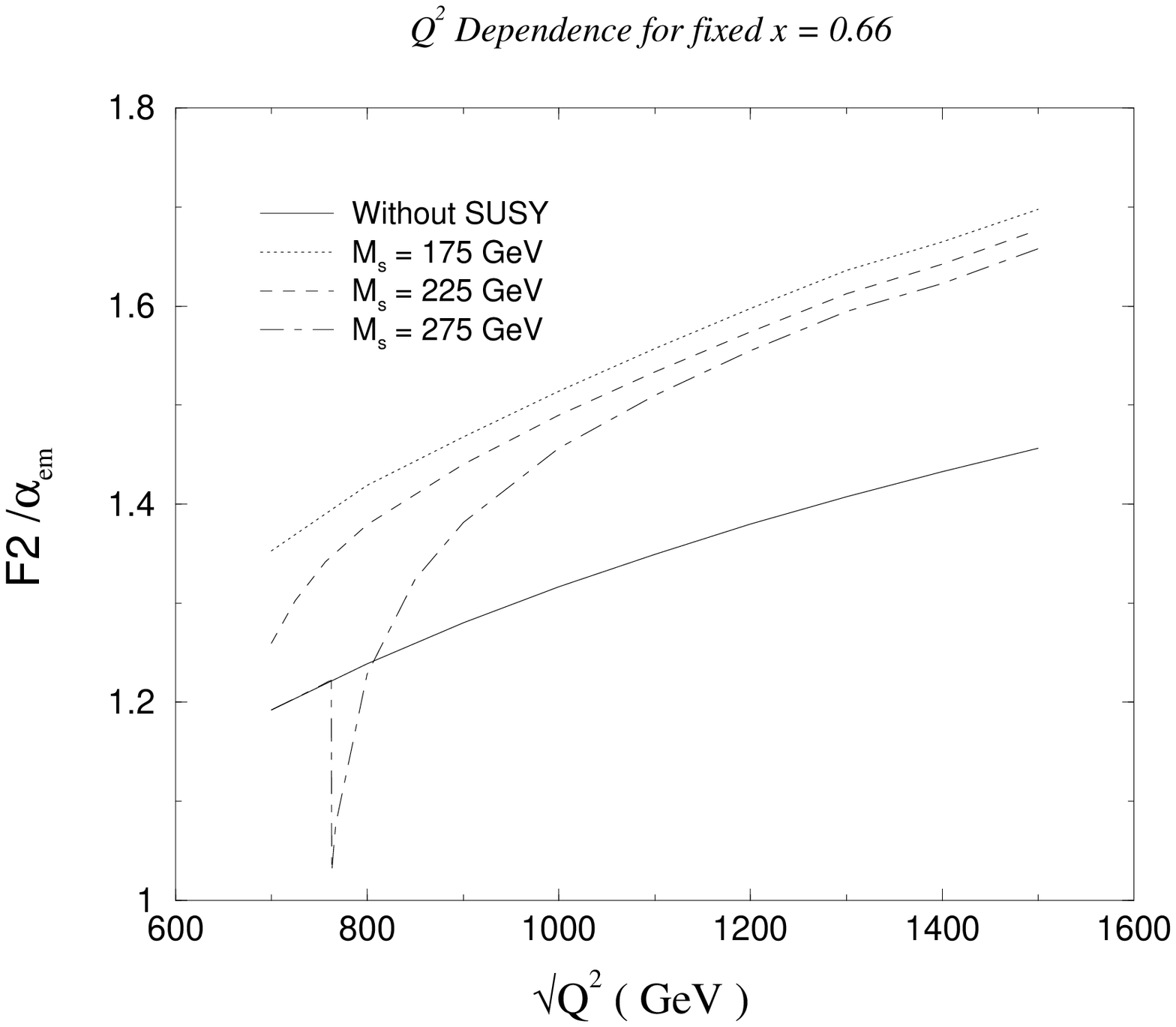}
\ecen
\caption{$x = 0.66$}
\label{f5}
\efig

\clearpage
\section{Conclusions}

We see from Figs. 1-5 that if one can build a machine for which values
of $Q^2$ approach $1 \  {\rm TeV}^2$
 (about twice the squark production threshold)
there is a substantial increase in the value of $\ftg$ for the photon.
Indeed, the evolution between the SUSY threshold and 1 TeV
is more than doubled if SUSY particles,  taken to have
a mass of 175 GeV, are present.  The difference between the
structure functions with and without SUSY in the middle range of Bjorken-$x$
is over 30\%, which should be easily detectable.

The effect  at $Q^2 =1 \  {\rm TeV}^2$ is, of course, diminished
if the SUSY threshold is increased. However, we note that taking the
squark masses to 300 GeV only has a small effect on $\ftg$. Conversely, if the
squark masses turn out to be substantially lighter than 175 GeV
 \footnote{The lowest value we take for the squark mass
is 175 GeV since this is above the threshold for t-quark and we find it
 useful to make comparisons of the evolution of the structure function
in the presence of SUSY with that without SUSY but with six active flavours.}
, (which is
not currently ruled out), there would be a significant effect on the
structure functions at values of $Q^2$ significantly
below $1\  {\rm TeV}^2$.

The effect also diminishes  if the target photon is off-shell, as will
usually be the case. However, we see from Figure 2 that
this effect is modest.

The results are fairly insensitive to the mass of the gluino.
This is not surprising as the gluino contributes very indirectly - it
can only be produced by a secondary emission from the target photon
 and then only probed through a further interaction with squarks.
Taking the mass of the gluino below that of the squark, would have
had a negligible effect as it is clear that it is the squark
threshold and not the gluino threshold that must be crossed before
there is any effect on the photon structure function.

In summary, we see that the effect of SUSY on the photon structure function
provides a further good reason for designing a large electron-positron
collider that would be capable of reaching values of $Q^2$ above the SUSY
threshold for the middle range of Bjorken-$x$.

\section*{Appendix A}

\def\theequation{A.\arabic{equation}}

The contribution to $F_2^{\gamma}$ of a left or right handed squark in deep
inelastic scattering on a photon has been calculated
\cite{TH3504},

\begin{eqnarray}
    F_{2,q}^{\gamma} \, & = & \, 3 e_q^4 \frac{\alpha}{\pi} x \Bigg\{
  \left[ 2x(1-x) + \tau x(3x-1) + \frac{1}{2} {\tau}^2 x^2  \right]
        \ln \left( \frac{1+v}{1-v}  \right) \nonumber \\
   & + &  \left[ 1-8x(1-x) + \tau x(1-x)  \right] v  \ \Bigg\} \label{L}
\end{eqnarray}

\noindent where,

\begin{eqnarray*}
\tau & = & 4M_s^2/Q^2 \\
 & & \\
v & = & [ 1 - \tau x/(1-x) ]^{1/2}
\end{eqnarray*}

We calculated this expression for the case $P^2 \neq 0$, where $r = P^2 / Q^2$.
The above relation is recovered on $r \to 0$ with,

\begin{eqnarray}
    F_{2,q}^{\gamma} \, & = & \, 3 e_q^4 \frac{\alpha}{\pi} x \Bigg\{
   B(M_s^2/Q^2)^2(1/FG) ( 16x^2 ) \nonumber \\
 & + & B(M_s^2/Q^2)(1/FG) (  - 48x^4r^2 + 48x^3r + 4x^2r - 8x^2 ) \nonumber \\
 & + & B(1/FG) (  - 12x^4r^3 + 12x^3r^2 - 2x^2r ) \nonumber \\
 & + & B ( 1 - 6x^2r + 6x^2 - 6x ) \nonumber \\
 & + & \ln(F/G)(M_s^2/Q^2)^2(B/\eta) ( 8x^2/b ) \nonumber \\
 & + & \ln(F/G)(M_s^2/Q^2)(B/\eta) ( 24x^4r^2/b + 2x^2r/b + 12x^2
          - 2x/b - 2x ) \nonumber \\
 & + & \ln(F/G)(B/\eta) \Big( \frac{1}{2} + 6x^4r^3/b + 12x^4r^2/b - 12x^3r^2/b
\nonumber \\
 & & - 12x^3r/b + 11x^2r/b - 3x^2r + 4x^2/b - 6x^2 + \frac{1}{2}xr/b \nonumber \\
 & & - \frac{1}{2}xr - 3x/b + 5x - \frac{1}{2b} \Big) \ \Bigg\}  \label{O}
\end{eqnarray}

\noindent where,

\begin{eqnarray*}
b & = & 1-2xr \\
 & &  \\
F & = & 1+\eta(1-2xr) \\
 & &  \\
G & = & 1-\eta(1-2xr) \\
 & &  \\
B & = & \sqrt{1 - \frac{4M_s^2x}{Q^2(1-x-xr)}} \\
 & &  \\
 & &  \\
\eta & = & \frac{B}{b} \sqrt{1-4x^2r}
\end{eqnarray*}

These equations are important in determining the $\gamma \to$ squark
splitting function i.e., the squark inhomogeneous $K_{S}$ and $K_{\Gamma}$ .
Also in determining the extra $S \! B_{\gamma}$ term in (Eq.~\ref{M}).
The choice is dependent on the squark threshold condition (Eq.~\ref{K}),
which is function of both $x$ and $Q^2$. At a particular $Q^2$ there will
be a region $x > x_t$ where squarks cannot be produced.


\end{document}